\documentclass{ws-p8-50x6-00}

\begin{document}

\title{QCD at Finite Density}

\author{Xiang-Qian Luo, Eric B. Gregory, Shuo-Hong Guo}

\address{Department of Physics, Zhongshan University
Guangzhou 510275, China}

\author{Helmut Kr\"oger}

\address{D\'epartement de Physique, Universit\'e Laval, Qu\'ebec, 
Qu\'ebec G1K 7P4, Canada}


\maketitle

\abstracts{At sufficiently high temperature and density, 
quantum chromodynamics (QCD) predicts 
phase transition 
from the hadronic phase to the quark-gluon plasma phase.
Lattice QCD is the most useful tool to investigate this
critical phenomenon, which status is briefly reviewed.
The usual problem in the Lagrangian formulation at finite density is
either an incorrect continuum limit or its complex action and
a premature onset of the
transition as the chemical potential is raised.
We show how the difficulties are overcome in our Hamiltonian approach.}

\section{Introduction}
\subsection{Motivation}

According to the standard model of cosmology,
12-15 billion years ago, the early universe 
underwent a series of drastic changes. 
Several microseconds after the big bang it was in a hot and dense 
quark-gluon plasma (QGP) state, 
where quarks and gluons were deconfined. Several minutes later, it changed to 
a phase, with the quarks and gluons confined inside the hadrons.
Today it remains in the low temperature 
and low density hadronic phase.
The ultimate goal of machines such as the Relativistic Heavy Ion Collider 
(RHIC) at BNL and the Large Hadron Collider (LHC) at CERN 
is to create the high temperature QGP phase 
and replay the birth and evolution of the universe.
In February 2000, physicists at CERN declared that they had experimentally 
found a new state of matter \cite{CERN}, which had never seen before.
Determining whether it is a QGP state requires further investigation 
from the future RHIC or LHC experiments.
QGP may also exist at low temperature,
in the core of very dense stars such as neutron stars.

Quantum chromodynamics (QCD) is the fundamental 
theory of quarks and gluons.
Figure 1 shows a recently proposed\cite{Raja} two-flavor QCD phase diagram  
on the temperature $T$ 
and chemical potential $\mu$ plane.
It assembles the information accumulated from lattice QCD, 
random matrix model,
and other approximation methods.
There is confinement and spontaneous 
chiral-symmetry breaking in the hadronic phase,
which is separated from other phases by a chiral phase transition.
The transition changes from second to
first order at a tricritical point. At high densities and low
temperatures, a superconducting phase ``2SC'' was recently found\cite{Raja} 
in which up and down quarks with two out 
of three colors form Cooper-pairs characterized by 
a diquark condensate $<\psi \psi>$.
The transition between this 2SC phase and the QGP phase 
is likely first order. The chiral phase transition on the horizontal
axis between the hadronic and 2SC phases is first 
order.

\begin{figure}[htb]
\centerline{\epsfxsize=9cm \epsfbox{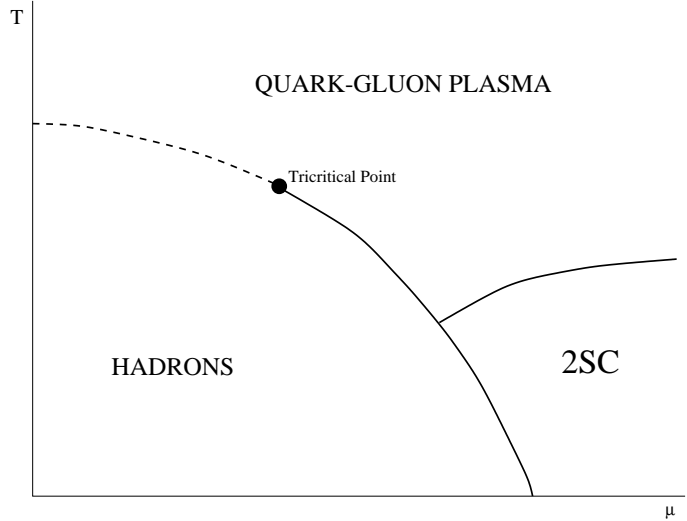}}
\caption{Phase diagram of QCD with 2 flavors.} 
\label{fig1}
\end{figure}

A precise understanding of the QCD phase structure
will  
provide valuable information in the  experimental search for the QGP.
Lattice gauge theory (LGT) 
proposed by Wilson in 1974, is a first principle nonperturbative 
technique for the investigation of phase transitions.
There are no free parameters
in LGT when the continuum limit is taken, 
in contrast with other nonperturbative techniques.
Although the standard lattice Lagrangian Monte 
Carlo method works very well 
for QCD at finite temperature,
it  unfortunately breaks down at non-zero chemical potential 
(due to the so-called complex action problem). This is briefly reviewed in
Sect. \ref{Status}.
In Sect. \ref{Our_Approach},
we present our recently developed Hamiltonian approach  \cite{QCD_mu} 
to lattice QCD at finite density, by introducing the
chemical potential in a natural way,
which avoids those usual problems.

\subsection{Present status}
\label{Status}

In the continuum, the grand canonical partition function of QCD 
at finite temperature $T$ and chemical potential $\mu$ is given by
\begin{equation}
Z=\rm{Tr} ~\rm{e}^{- \beta \left( H - \mu N \right)}, ~~~ 
\beta = (k_{B} T)^{-1} , 
\label{partition}
\end{equation}
where $k_B$ is the Boltzmann constant, $H$ is the Hamiltonian, 
and $N$ is particle number operator
\begin{equation}
N=\int d^3x ~ \psi^{\dagger}(x) \psi (x).
\label{PartNumber}
\end{equation}
The energy density of the system with free quarks is given by \cite{Li}
\begin{equation}
\epsilon = {1 \over V}{1 \over Z} \rm{Tr} ~ H ~ \rm{e}^{- \beta \left( H- \mu N \right)}
= -{1 \over V} \frac{\partial \ln Z} { \partial \beta } 
\vert_{\mu \beta} .
\end{equation} 
Going over to $T \to 0$ and chiral limit $m \to 0$, one gets the energy density 
(where the contribution of $\mu=0$ is subtracted)
\begin{equation}
\epsilon_{sub} =
{\mu^4 \over 4 \pi^2}.
\label{cont}
\end{equation}
Unfortunately, in Lagrangian LGT, a naive discretization 
of the chemical potential term does not lead
to the correct continuum relation Eq.(\ref{cont}).  
Let us take the naive fermions as an example.
The action reads 
\begin{eqnarray}
S_f= a^4 \sum_x  m \bar{\psi}(x) \psi(x) 
+ {a^3 \over 2} \sum_{x,k} 
\bar{\psi}(x) \gamma_k \psi(x+\hat{k})
+ a^4 \mu \sum_x \psi^{\dagger}(x) \psi (x),
\label{naive}
\end{eqnarray}
where $\gamma_{-k}=-\gamma_{k}$. 
In the chiral limit $m \to 0$ and the continuum limit $a \to 0$, 
the subtracted energy density from this action
is
$\epsilon_{sub} \propto (\mu / a)^2$, i.e. becoming quadratically divergent \cite{Hasenfratz}, 
and therefore it is inconsistent with the continuum result of Eq.(\ref{cont}).
This problem is not due to the species doubling of naive fermions, because
the case of Kogut-Susskind fermions or Wilson fermions is similar.

Hasenfratz and Karsch \cite{Hasenfratz} proposed to
introducing the chemical potential 
exponentially
\begin{eqnarray}
S_f &=& a^4 \sum_x  m \bar{\psi}(x) \psi(x) 
+ {a^3 \over 2} \sum_{x} \sum_{j=1}^3 
\left[ \bar{\psi}(x) \gamma_j \psi(x+\hat{j})
-\bar{\psi}(x+\hat{j}) \gamma_j \psi(x) \right] 
\nonumber \\
&+& {a^3 \over 2} \sum_x  
\left[ e^{\mu a} \bar{\psi}(x) \gamma_4 \psi (x+\hat{4})
-e^{-\mu a} \bar{\psi}(x+\hat{4}) \gamma_4 \psi (x)\right].
\label{Hasen}
\end{eqnarray}
The chemical potential can be introduced analogously for KS 
as well as for Wilson fermions.
Such treatment of the chemical potential is numerically feasible 
in the quenched approximation (where the fermionic determinant $\det \Delta$ 
is constraint to be $1$, and quark loops are suppressed).
However, there is evidence \cite{Quenched} that the quenched approximation
produces an unphysical onset of the critical chemical potential at the value
$\mu_{C}=M_{\pi} (m \not= 0)/2$, which goes to zero in the chiral limit
[$M_{\pi}(m \not=0)$ is the pion mass at finite bare quark mass $m$. 
A finite bare quark mass has to be introduced 
in most of the numerical simulations].
This is in conflict with other theoretical predictions $\mu_C \approx M_N^{(0)}/3$ 
[$M_N^{(0)}$ is the nucleon mass at $\mu=0$].

For full QCD, the fermionic degrees of freedom 
have to be integrated out, and  
in the measure occurs  $\det \Delta$.
For finite chemical potential $\det \Delta$ from Eq. (\ref{Hasen})
and the effective fermionic action $\ln \det \Delta$  become complex, 
due to $\Delta_{y,x} \not= \gamma_5 \Delta_{x,y}^{\dagger} \gamma_5$, 
which renders numerical simulations
extremely difficult. 
Much effort has been made to solve this notorious complex
action problem: \\
\noindent (1) The Glasgow group has suggested to treat 
$\det \Delta$ as observable\cite{Glasgow}.
This method requires a very large number of configurations, 
in particular for $\mu \approx \mu_C$. Even on a very small lattice $V=4^4$, 
the computational costs exceed the current computer capacity \cite{Zaragoza}. 
Even in this case,
the unphysical onset of $\mu_C$ still exists.
Therefore, it is unclear  whether the onset is an intrinsic problem of
the proposal Eq. (\ref{Hasen}).\\
\noindent (2) In the imaginary chemical potential 
method \cite{Imaginary} $\det \Delta$ becomes real, which works well  
for numerical simulations at high temperature and low density. 
But it might not work at low temperature and high density. \\
\noindent (3) It has been proposed to utilize a special symmetry \cite{Hands}. 
This is the only successful method in Lagrangian formulation, 
and attracts much attention \cite{Zaragoza_qq,Liu},
but it works only for the SU(2) gauge group. \\
\noindent (4) In \cite{Zaragoza_qq},  the probability
distribution function method \cite{pdf} 
is applied to 2-color QCD (free of complex
action problem), and
very interesting results for the diquark
condensate $<\psi \psi>$ are obtained.

\section{Hamiltonian Approach}
\label{Our_Approach}
\subsection{Free fermions at zero chemical potential}

The lattice Hamiltonian describing noninteracting Wilson fermions 
in $d+1$ dimensions at $\mu=0$ is
\begin{equation}
H= \sum_x  m \bar{\psi}(x) \psi(x) 
+ \sum_{x,k}{1 \over 2a} \bar{\psi}(x) \left( \gamma_k \psi(x+\hat{k}) 
+ r \left(\psi (x) - \psi (x+\hat{k})\right)\right).
\label{Hamiltonian_mu0}
\end{equation}
The up and down components of $\psi$ are coupled via the $\gamma_k$ matrices, 
and one can use a unitary transformation  
\cite{Luo}
\begin{eqnarray}
H'= \exp(-iS) ~ H  ~ \exp(iS)
\end{eqnarray}
to decouple them.
The operator $S$ can be computed explicitly \cite{Luo} in momentum space:
\begin{eqnarray}
S = -\sum_p {\theta_p \over A_{p}}
\sum_{j=1}^d \psi^{\dagger}_p \gamma_{j} \psi_p {\sin p_j a \over a},
~~~ 
A_{p} = \left( \sum_{j=1}^d \left({\sin p_j a \over a}\right)^2 \right)^{1/2}.
\label{operator}
\end{eqnarray}
The physical vacuum state of $H$ reads
\begin{equation}
\vert \Omega \rangle = \exp(iS) \vert 0 \rangle,
\end{equation}
where $\vert 0 >$ is the bare vacuum state defined as 
$\xi \vert 0\rangle = \eta \vert 0\rangle =0$.
The parameter $\theta_p$ in Eq. (\ref{operator}) is so determined that
the vacuum energy
$E_{\Omega} =
\langle \Omega \vert H \vert \Omega \rangle
= \langle 0 \vert H' \vert 0 \rangle$
is minimized.
After the unitary transformation, the 
fermionic field $\psi$ can be simply expressed by  
up and down 2-spinors $\xi$ and $\eta^{\dagger}$ as 
$\psi =\left( 
\begin{array}{c}
\xi \\
\eta^{\dagger} 
\end{array}
\right)$ 
and  $H'$ is diagonal
\begin{eqnarray}
H'  =  
\sum_{p} 
A'_p
\bar{\psi}_p \psi_p
= \sum_p 
\left( \left( m+ 
{2r \over a} \sum_{j=1}^d \sin^2 \left( p_j a/2 \right) \right)^{2}
+ A_{p}^{2} \right)^{1/2} \bar{\psi}_p \psi_p. 
\label{H'}
\end{eqnarray}
For Wilson fermions, 
in the continuum limit $a \to 0$, for any finite momentum $p$, we have
$E_{\Omega}=-2N_{c}N_{f}\sum_{p} \sqrt{m^2+p^2}$,
giving the correct dispersion relation.
Here $N_c$ and $N_f$, respectively, are the number of colors 
and number of flavors.

\subsection{Free fermions at nonzero chemical potential}
\label{free_fermion_mu}
Now we naturally introduce the chemical potential
\begin{eqnarray}
H_{\mu}=H- \mu N,
\label{Hamiltonian}
\end{eqnarray}
where $H$ is given by Eq.(\ref{Hamiltonian_mu0}) and $N$ is given by 
Eq.(\ref{PartNumber}).
Let us define the state
$\vert n_p, \bar{n}_p  \rangle$ by
\begin{eqnarray*}
&& \xi_p \vert 0_p, \bar{n}_p  \rangle=0,
~~ \xi^{\dagger}_p \vert 0_p, \bar{n}_p  \rangle=\vert 1_p, 
\bar{n}_p  \rangle, ~~~ \xi_p \vert 1_p, 
\bar{n}_p  \rangle=\vert 0_p, \bar{n}_p  \rangle,
~~ \xi^{\dagger}_p \vert 1_p, \bar{n}_p  \rangle=0, 
\nonumber \\
&& \eta_p \vert n_p, 0_p  \rangle=0,
~~ \eta^{\dagger}_p \vert n_p, 0_p  \rangle=\vert n_p, 1_p  \rangle, 
~~ \eta_p \vert n_p, 1_p  \rangle=\vert n_p, 0_p  \rangle,
~~ \eta^{\dagger}_p \vert n_p, 1_p  \rangle=0 .
\label{state}
\end{eqnarray*}
The numbers $n_p$ and $\bar{n}_p$ take the values 0 
or 1 due to the Pauli principle. 
By definition, the up and down components of the fermion field
are decoupled. Obviously, this is not an eigenstate of $H_{\mu}$ due to
the non-diagonal form of $H$.
However, they are eigenstates of $H'_{\mu}$, 
which are related to $H_{\mu}$ by a unitary transformation
\begin{equation}
H_{\mu}'= \exp(-iS) ~ H_{\mu}  ~ \exp(iS)=H'- \mu N.
\label{unitary_mu}
\end{equation}
For the vacuum eigenstate of $H_{\mu}$ we make an ansatz of the following form
\begin{equation}
\vert \Omega \rangle =  \exp(iS)\sum_p f_{n_p, \bar{n}_p} 
\vert n_p, \bar{n}_p  \rangle.
\end{equation}
$S$ is given by Eq.(\ref{operator}), 
and $H'$ is given by Eq. (\ref{H'}).
The vacuum energy is
$E_{\Omega} =
2N_c N_f 
\sum_{p} C_{n_p, \bar{n}_p} \left[ \left( A'_p-\mu \right) n_p  
+ \left( A'_p+ \mu \right)\bar{n}_p - A'_p- \mu \right]$.
Here we have introduced the notation $C_{n_p, \bar{n}_p}=f_{n_p, \bar{n}_p}^2$,
which have not yet been specified.
For this purpose we use the stability condition of the vacuum.   
Because $\mu >0$, the vacuum energy increases with $n_p$.
This means the vacuum is unstable unless $\bar{n}_p=0$.
This simplifies the vacuum energy  to
$E_{\Omega} =
2N_c N_f 
\sum_{p} \left[ C_{1_p}\left( A'_p-\mu \right)  
- A'_p - \mu \right]$,
where we use the abbreviation $C_{n_p}=C_{n_p,0}$ and the
normalization condition.
The dependence of $C_{1_p}$ on the value of $\mu$ 
can be seen by inspection of the derivative
$\partial E_{\Omega}/ \partial C_{1_p}
= 2N_c N_f \left( A'_p-\mu \right)$.
For $\mu > A'_p$, the right-hand side 
is negative. Maximizing $C_{1_p}$ means minimizing the vacuum energy. 
Therefore, $C_{1_p}=1$.
On the other hand, for $\mu < A'_p$, the right-hand side 
is positive and for any $C_{1_p}$ the vacuum is unstable. 
Therefore, $C_{1_p}=0$. We can summarize these properties by writing
$C_{1_p} = \Theta \left( \mu -A'_p\right)$.
Thus the vacuum expectation 
\footnote{There is a typo in Eqs. (2.24) and (2.25) 
in \cite{QCD_mu}:
$E_{\Omega}$ there should be read as 
$\langle \Omega \vert H \vert \Omega \rangle$.
The results and conlusions are the same.}
of $H$ becomes 
$\langle \Omega \vert H \vert \Omega \rangle = 2N_c N_f 
\sum_{p} \left( C_{1_p} A'_p  - A'_p \right)$.
The subtracted energy density reads
\begin{equation}
\epsilon_{sub} = \frac{ \langle \Omega \vert H \vert \Omega \rangle
 - \langle \Omega \vert H \vert \Omega \rangle\vert_{\mu=0} }
{ N_c N_f N_s }
={\mu^4 \over 4 \pi^2}.
\label{lat_hamil}
\end{equation}
Here $N_s$ is the number of spatial lattice sites.
Thus we have proven that our Hamiltonian approach to 
free quarks at finite chemical potential leads to 
the correct continuum result for the vacuum energy density, 
Eq. (\ref{cont}).
From this relation, we can easily
see that the free quark number density is proportional to $\mu^3$.
For naive fermions, in the continuum limit $a=0$, 
there will be an extra factor of $2^d$.

\subsection{Strong coupling QCD at nonzero chemical potential}

As is well known, lattice QCD at $\mu=0$ confines
quarks and spontaneously breaks chiral symmetry.
For a sufficiently large chemical potential, this picture may change.
Here we set out to investigate finite density QCD 
in the strong coupling regime $1/g^2 << 1$. 
Following reference\cite{Luo},
$H'$ in Eq. (\ref{unitary_mu}) is now replaced by
\begin{eqnarray}
H' &=& \left[ m\left[ 1-\left( 2 \theta_0 \right)^2 d \right] 
\hspace{-0.1cm} + \hspace{-0.1cm} 
{\left( 2 \theta_0 \right)d \over a} \right] \sum_x \bar{\psi}(x) \psi(x)
+ {g^2 C_N d \left( 2 \theta_0 \right)^2 \over 4a}\sum_x \psi^{\dagger}(x) \psi(x) 
\nonumber \\
&-& {g^2 C_N \left( 2 \theta_0 \right)^2 \over 32aN_c} 
\sum_{x,k} L_A \psi^{\dagger}_{f_1}(x) \Gamma_A \psi_{f_2}(x)
\psi^{\dagger}_{f_2}(x+k) \Gamma_A \psi_{f_1}(x+k). 
\hspace{-0.5cm}
\label{Hamiltonian_fierz}
\end{eqnarray}
The additional four-fermion term 
is induced by gauge interactions with fermions.
There $d=3$ denotes the spatial dimension, 
$f_1, ~f_2$ are flavor indices 
(summation over repeated indices is understood),
$\theta_0=1/(4ma+g^2C_N)$, and $C_N=(N_c^2-1)/(2N_c)$.
The matrices $\Gamma_A$ and their coefficients $L_A$ are given in Tab.[1].

\begin{table}
\begin{center}
\begin{tabular}{|c|c|c|c|c|c|c|c|c|}
\hline
$\Gamma_A$ &\hspace{-0.1cm} 1\hspace{-0.1cm} & 
$\hspace{-0.1cm} \gamma_j \hspace{-0.1cm}$ & 
$\hspace{-0.1cm}\gamma_4 \hspace{-0.1cm}$ & 
$\hspace{-0.1cm}\gamma_5\hspace{-0.1cm}$ 
& $\hspace{-0.1cm} i\gamma_4 \gamma_5\hspace{-0.1cm}$ 
& $\hspace{-0.1cm} i\gamma_4 \gamma_j \hspace{-0.1cm}$ 
&$\hspace{-0.1cm} i \epsilon_{j j_1 j_2} \gamma_{j_1} \gamma_{j_2}
\hspace{-0.1cm}$ 
& $\hspace{-0.1cm}i\epsilon_{j j_1 j_2}\gamma_4\gamma_{j_1}\gamma_{j_2}\hspace{-0.1cm}$\\
\hline
$L_A$ &\hspace{-0.1cm} 1\hspace{-0.1cm} 
& $ \hspace{-0.1cm} 2\delta_{k,j}-1 \hspace{-0.1cm}$ 
&\hspace{-0.1cm} -1\hspace{-0.1cm} 
&\hspace{-0.1cm} -1\hspace{-0.1cm} 
&\hspace{-0.1cm} 1\hspace{-0.1cm} 
& $\hspace{-0.1cm}1-2\delta_{k,j}\hspace{-0.1cm}$ 
&$\hspace{-0.1cm} 2\delta_{k,j}-1 \hspace{-0.1cm}$ 
&$\hspace{-0.1cm}1-2\delta_{k,j}\hspace{-0.1cm}$\\
\hline
\end{tabular}
\end{center}
\caption{$\Gamma$ matrices and coefficients.}
\end{table}

The unitary transformed Hamiltonian  Eq. (\ref{Hamiltonian_fierz})
can also be re-expressed 
in terms of the following pseudoscalar and vector operators \cite{Luo}
\begin{eqnarray}
&& 
\Pi ={1 \over 2 \sqrt {- \bar{v} }}
\psi^{\dagger}\left(1-\gamma_4 \right) \gamma_5  \psi, 
\nonumber \\
&& 
V_j={1 \over 2 \sqrt {- \bar{v}}}
\psi^{\dagger}\left(1-\gamma_4 \right)\gamma_j \psi, 
\label{mesons}
\end{eqnarray}
as \cite{QCD_mu}
\begin{eqnarray}
H' &=& E_{\Omega}^{(0)}+G_1 \sum_p 
\left( \Pi^{\dagger}(p) \Pi(p) + \sum_j V_j^{\dagger}(p) V_j(p) \right)
\nonumber \\
&+& G_2 \sum_{p}\left( \Pi^{\dagger}(p) \Pi^{\dagger}(-p) +h.c. \right) 
\sum_j \cos p_j a 
\nonumber \\
&+& G_2 \sum_{p,j} \left( V_j^{\dagger}(p) V_j^{\dagger}(-p) + h.c. \right)  
\left(\sum_{j'} \cos p_{j'} a - 2 \cos p_j a \right),
\label{Hamiltonian_mesons_for}
\end{eqnarray}
where
\begin{eqnarray*}
E_{\Omega}^{(0)} &=& N_f N_s
\left[ m\left[1-\left( 2 \theta_0 \right)^2 d \right] 
+{\left( 2 \theta_0 \right)d \over a} \right]  \bar{v} 
\nonumber \\
&+& N_f N_s {g^2 C_N d \left( 2 \theta_0 \right)^2 \over 4a} v^{\dagger} 
- N_f N_s {g^2 C_N \left( 2 \theta_0 \right)^2 d  \over 16aN_c}  
\left( v_2^{\dagger}- \bar{v}_2 \right),
\nonumber \\
G_1 &=&
2 \left[ m\left[1-\left( 2 \theta_0 \right)^2 d \right] 
+{\left( 2 \theta_0 \right)d \over a} \right] 
+ {g^2 C_N d \left( 2 \theta_0 \right)^2 \over 4 a N_c} \bar{v},
\nonumber \\
G_2 &=& - {g^2 C_N \left( 2 \theta_0 \right)^2 \over 8 a N_c}  \bar{v},
\nonumber \\
v^{\dagger} &=&
{2N_c \over N_s}
\sum_{p} C_{n_p, \bar{n}_p} \left( n_p - \bar{n}_p + 1 \right),
\nonumber \\
\bar{v}&=&
{2N_c \over N_s}
\sum_{p} C_{n_p, \bar{n}_p} \left( n_p + \bar{n}_p -1 \right),
\nonumber \\
v_2^{\dagger} &=&
{(2N_c)^2  \over N_s}
\sum_{p} C_{n_p, \bar{n}_p} \left( n_p - \bar{n}_p + 1 \right)^2,
\nonumber \\
\bar{v}_2 &=&
{(2N_c)^2  \over N_s}
\sum_{p} C_{n_p, \bar{n}_p} \left( n_p + \bar{n}_p -1 \right)^2.
\end{eqnarray*}
Eq. (\ref{Hamiltonian_mesons_for}) 
can be diagonalized by a Bogoliubov transformation \cite{Luo}
\begin{eqnarray*}
P(p) &=& \cosh u(p) a(p) + \sinh u(p) a^{\dagger}(-p),
\nonumber \\
V_j(p) &=& \cosh v_j(p) b(p) + \sinh v_j(p) b^{\dagger}(-p),
\end{eqnarray*}
\begin{eqnarray*}
\tanh 2u(p) &=& {-2G_2 \over G_1}\sum_j \cos p_j a,
\nonumber \\
\tanh 2v_j(p) &=& {-2G_2 \over G_1}\left(\sum_{j'} \cos p_{j'} a
-2 \cos p_j a \right).
\end{eqnarray*}
The Bogoliubov transformed Hamiltonian eventually becomes
\begin{eqnarray}
H'' & = & E_{\Omega}^{(0)}
+ {N_f^2 \over 2} G_1 \sum_p \left[ \sqrt{1-\tanh^2 2u(p)} -1 \right]
\nonumber \\
&+& {N_f^2 \over 2} G_1 \sum_{p,j} \left[ \sqrt{1-\tanh^2 2v_j(p)} -1 \right]
\nonumber \\
&+& G_1 \sum_p \sqrt{1-\tanh^2 2u(p)} a^{\dagger}(p) a(p)
\nonumber \\
&+& G_1 \sum_{p,j} \sqrt{1-\tanh^2 2v_j(p)} b_j^{\dagger}(p) b_j(p) .
\label{Bogoliubov_H}
\end{eqnarray}
Using the notation, normalization condition and arguments 
for the coefficient as in
Sect. \ref{free_fermion_mu}, we obtain $\bar{n}_p=0$ and
%
$C_{1_p} = \Theta \left( \mu -m_{dyn}^{(0)}\right)$.
%
Here $m_{dyn}^{(0)}=d / (ag^2C_N)$ is the dynamical quark mass at $\mu=0$.
Therefore, in the chiral limit $m=0$, the vacuum energy is
\begin{eqnarray}
\frac{ E_{\Omega} }{ 2 N_c N_f N_s} & =&
\left(m_{dyn}^{(0)}-\mu \right) \Theta \left(\mu -m_{dyn}^{(0)}\right)
-m_{dyn}^{(0)}-\mu 
\nonumber \\
&+&  {N_f \over 2 N_s} G_1 \sum_p \left[ \sqrt{1-\tanh^2 2u(p)} -1 \right]
\nonumber \\
&+& {N_f \over 2 N_s} G_1 \sum_{p,j} \left[ \sqrt{1-\tanh^2 2v_j(p)}-1 \right].
\label{result1} 
\end{eqnarray}

According to the Feynman-Hellmann theorem,
the chiral condensate is related to the ground state energy by
\begin{eqnarray}
\langle \bar{\psi} \psi \rangle =
{1 \over N_f N_s} \lim_{m\to 0} 
{\partial E_{\Omega} \left( m\not=0 \right)
\over \partial m}
= \langle \bar{\psi} \psi \rangle^{(0)}
\left[ 1-\Theta \left( \mu- m_{dyn}^{(0)}\right) \right] ,
\label{result2} 
\end{eqnarray}
where $\langle \bar{\psi} \psi \rangle^{(0)}$
is the chiral
condensate at $\mu=0$
\begin{eqnarray}
\langle \bar{\psi} \psi \rangle^{(0)}
=-2N_C \left( 1 - {4d \over g^4 C_N^2} \right) \left(1-{N_f \over N_c} I_1
-{N_f \over N_c} I_2 \right) 
\end{eqnarray}
and for $d=3$, $I_1=0.078354 \pm 2 \times 10^{-6}$, $I_2 =   0.235075 \pm 4 \times 10^{-6}$.
According to Eq. (\ref{result2}), for $\mu < m_{dyn}^{(0)}$, 
$\langle \bar{\psi} \psi \rangle
=\langle \bar{\psi} \psi \rangle^{(0)} \not=0$, i.e.,
chiral symmetry is spontaneously broken.
For $\mu > m_{dyn}^{(0)}$, $\langle \bar{\psi} \psi \rangle=0$,
i.e., chiral symmetry is restored. Therefore, there is a first order
chiral phase transition and the critical value of $\mu$ is given by
\begin{equation}
\mu_C = m_{dyn}^{(0)} = {d \over g^2 C_N a},
\label{result3} 
\end{equation}
The critical chemical potential $\mu_C$ is equal 
to the dynamical quark mass at $\mu=0$, 
which agrees with the result from an entirely different method \cite{Yaouanc88_1} in Hamiltonian formulation.
(The authors argued this was a second order
phase transition, in contrast we clearly observe a first order transition).
Our result is consistent with other theoretical predictions 
$\mu_C \approx M_N^{(0)}/3$, 
because (see below) at $\mu=0$ holds $M_N^{(0)} \approx 3m_{dyn}^{(0)}$.

We can compute now the quark number density in the chiral limit $m=0$, which yields
\begin{equation}
n_q = {-1 \over 2 N_c N_f N_s}
{\partial E_{\Omega} 
\over \partial \mu}-1
=
\frac{ \langle \Omega \vert \sum_{x} \psi^{\dagger}(x) \psi (x) \vert \Omega \rangle } 
{2 N_c N_f N_s }-1
= \Theta \left(\mu-\mu_C \right),
\label{result4} 
\end{equation}
This is consistent with the $\beta=0 $ simulation results 
described in \cite{FirstOrder}, 
and however,  is different from 
the large $\mu$ behavior  in the continuum ({\it i.e.} 
the Stefann-Boltzmann law $n_q \propto \mu^3$ for free quarks).
It remains to be seen whether higher order $1/g^2$ calculations
will improve this behavior.

The quark number susceptibility, standing for the response of the 
quark number density to infinitesimal changes in $\mu$, is
\begin{equation}
\chi_q = 
\frac{ \partial n_q} 
{\partial \mu }
= \delta \left(\mu-\mu_C \right).
\label{result5} 
\end{equation}

Finally, let us look at some implications on the thermal mass spectrum
of the pseudoscalar meson, vector meson and nucleon.
The thermal mass is defined by
$M_{h}^{\star} = \langle h \vert H- \mu N \vert h \rangle -E_{\Omega}$.
For the pseudoscalar meson, in the chiral limit $m=0$,
\begin{eqnarray}
M_{\pi}^{\star}
 &= &G_1 \sqrt{1-\tanh^2 2u(p=0)}
=\left\{
\begin{array}{cc}
0 & ~~~ \rm{for} ~~~ \mu < \mu_C, \\
4m^{(0)}_{dyn} &~~~ \rm{for} ~~~ \mu > \mu_C . 
\end{array}
\right .
\label{result6} 
\end{eqnarray}
Therefore, in the broken phase,
the pseudoscalar is a Goldstone boson ($M_{\pi}^{\star} \propto \sqrt{m} \to 0$), 
and in the symmetric phase,
it is no longer a Goldstone boson. For the vector meson,
\begin{eqnarray}
M_V^{\star}
 &= &G_1 \sqrt{1-\tanh^2 2v_j(p=0)}
=\left\{
\begin{array}{cc}
M_V^{(0)} & ~~~ \rm{for} ~~~ \mu < \mu_C, \\
4m^{(0)}_{dyn} &~~~ \rm{for} ~~~ \mu > \mu_C . 
\end{array}
\right .
\label{result7} 
\end{eqnarray}
where $M_V^{(0)}=4 \sqrt{d-1}/(ag^2C_N)$ is the vector mass
at $\mu=0$. Therefore, $\partial M/\partial \mu 
\propto \delta(\mu-\mu_C)$ for the pseudoscalar and vector mesons. 
To see the critical behavior at zero temperature, 
one should be very close to $\mu_C$.
This behavior is consistent with that of the quark number density.
To see whether the meson thermal masses depend on $\mu$, 
higher order $1/g^2$ corrections must be included.

For the nucleon, we obtain the expected behavior
\begin{eqnarray}
M_N^{\star}
 &= & 
M_N^{(0)}- 3\mu
\label{result8}
\end{eqnarray}
for  $\mu < \mu_C$,
where $M_N^{(0)} \approx 3m_{dyn}^{(0)}$.
This leads to 
$M_N^{\star}=0$ at $\mu = \mu_C$.

\section{Outlook}
In this paper, we have developed a Hamiltonian approach 
to lattice QCD at finite density. 
The chemical potential is introduced in a very natural way
as in the continuum.
It avoids the persisting problem in the Lagrangian approach
of an incorrect continuum limit, or complex action or
a premature onset of the
transition to nonzero quark density as $\mu$ is raised.
The main result in the free case
is given by Eq. (\ref{lat_hamil}), and those in the strong coupling
regime are given by Eqs. (\ref{result1})-(\ref{result8}).
We have seen that the approach
works well in the free case and also in the strong coupling regime. 
We predict that at strong coupling, the chiral transition is of first order,
and the critical chemical potential $\mu_C \approx M_N^{(0)}/3$.

We have not yet specified the nature of the
chiral-symmetric phase for $\mu > \mu_C$.
Is it a QGP phase or a color-superconduction phase \cite{Raja} (see Fig. 1)?
Up to now, there has been no first principle investigation of such a phase.
Better understanding of it will affect 
our knowledge of the formation of the neutron star.

At this point, we should mention that the results of Hamiltonian
formulation and Lagrangian formulation \cite{Lagrangian_strong}
might be quite different.
It is interesting to see whether they describe the same physics
in the continuum.
For pure gauge theory, within a Hamiltonian approach,
we can extend to the intermediate coupling 
and obtain meaningful results for the glueballs \cite{glueball}. 
For fermions, the calculation is far from trivial.
Recently we proposed a Monte Carlo technique 
in the Hamiltonian formulation \cite{mch} for the purpose to do 
nonperturbative numerical simulations, by combining the virtues of 
the Monte Carlo algorithm with importance sampling 
and the Hamiltonian approach.
We hope to apply it to QCD and with the aim 
to obtain useful information for RHIC and LHC physics and neutron star phenomenology.

\bigskip

\noindent
{\bf Acknowledgments}

We are grateful to the attendees of the workshop for useful
discussions.
X.Q.L. is supported by the
National Science Fund for Distinguished Young Scholars (19825117),
National Natural Science Foundation (10010310687),
the Ministry of Education,
the Doctoral Program of Higher Education and   
Guangdong Provincial Natural Science Foundation (990212) of China.
X.Q.L. and E.B.G. are supported by 
the Guangdong National Communication Ltd.
H.K. has been supported by NSERC Canada.

\end{document}